\documentclass{article}
\usepackage{spconf,graphicx}
\usepackage{epsfig}
\usepackage{amsmath,amssymb,amsthm,amsfonts}

\newcommand{\bqn}{\begin{eqnarray}}
	\newcommand{\eqn}{\end{eqnarray}}
\newcommand{\bq}{\begin{eqnarray*}}
	\newcommand{\eq}{\end{eqnarray*}}

\newcommand{\ba}{\left( \begin{array}}
\newcommand{\ea}{\end{array} \right)}

\usepackage{url}
\usepackage{hyperref}
\hypersetup{colorlinks,%
citecolor=black,%
filecolor=black,%
linkcolor=red,%
urlcolor=black,%
pdftex}

\usepackage{xcolor}

\graphicspath{{figures/}}

\title{Poisson Flow Model of Cortical Folding Pattern}
\name{
Moo K. Chung$^1$\thanks{Published in IEEE Engineering in Medicine and Biology Society Annual Conference (EMBC) 2026. The correspondence should be sent to {\tt mkchung@wisc.edu}. This work was supported by NIH MH133614 and NSF DMS--2010778.}
Luigi Maccotta$^2$, %l.maccotta@wustl.edu
Aaron Struck$^2$ %struck@wustl.edu
}
\address{$^1$ University of Wisconsin, Madison, USA\\
		$^2$Washington University, St. Louis, USA}

\begin{document}
\maketitle

\begin{abstract}
Cortical folding reflects coordinated neurodevelopmental processes and provides a sensitive marker of neurological disease. In juvenile myoclonic epilepsy (JME), structural abnormalities are subtle and spatially distributed, limiting the sensitivity of conventional morphometric measures such as cortical thickness. 
We introduce a Poisson flow model derived from gradients of the mean curvature field on the cortical surface. The method yields a smooth scalar field obtained from a Poisson equation, whose surface gradient defines a flow representation of folding organization. This representation enables spatially coherent characterization of sulcal--gyral patterns and provides a principled geometric framework for studying distributed cortical alterations in JME.
\end{abstract}

\section{Introduction}

Cortical folding reflects a complex interplay between neurodevelopmental processes, mechanical constraints, and differential growth, and its organization is closely linked to normal brain function and neurological disease \cite{vanessen.1997,zilles.2013}. Accordingly, abnormalities in sulcal–gyral morphology have been reported across a range of conditions, including epilepsy \cite{wandschneider.2019}, neurodevelopmental disorders \cite{chung.2005.NI}, and neurodegenerative diseases \cite{kim.2014.NI}, indicating that folding patterns encode clinically meaningful information beyond gross cortical shape.

Juvenile myoclonic epilepsy (JME) is a genetic epilepsy with adolescent onset and a strong neurodevelopmental component \cite{wandschneider.2019,struck.2025}. Prior morphometric studies have reported abnormalities involving fronto-central and higher-order association regions \cite{wandschneider.2019}, while our recent volumetric work demonstrates selective alterations in motor-associated thalamic nuclei and corticothalamic circuits \cite{struck.2025}. These findings support the view that JME represents a distributed neurodevelopmental  disorder rather than focal cortical pathology.

Despite this recognition, most cortical morphometry studies rely on indirect summary measures such as cortical thickness, gyrification index, or regional volumes. These scalar descriptors primarily quantify local gray-matter properties and provide limited information about the intrinsic geometry and spatial organization of sulcal–gyral folding patterns \cite{huang.2020.TMI,im.2019}. Although effective for detecting regional differences, such measures collapse complex folding architecture into isolated local quantities and therefore cannot capture coordinated folding organization across the cortical surface. This limitation is particularly consequential in JME, where abnormalities are thought to reflect large-scale neurodevelopmental alterations spanning distributed brain regions rather than focal cortical changes.

To address this gap, we model cortical folding using a Poisson equation applied to the mean curvature field on the cortical surface. Solving the Poisson equation yields a smooth scalar field, whose gradient captures transitions from regions of higher to lower curvature. This yields a spatially coherent representation that preserves relative geometric organization while suppressing high-frequency noise. The resulting representation enables statistically efficient, vertexwise inference on folding organization and provides a principled geometric alternative to conventional morphometric measures for studying distributed neurodevelopmental alterations in JME.

\section{Materials and Methods}

\subsection{Imaging Data and Preprocessing}

Participants were drawn from the Juvenile Myoclonic Epilepsy Connectome Project (JMECP), a prospective study conducted at the University of Wisconsin Hospital and Clinics \cite{struck.2025}. Detailed screening procedures and inclusion and exclusion criteria are reported in \cite{struck.2025}. The present study included 61 individuals with JME (mean age 19.72 $\pm$ 3.70 years) and 39 healthy controls (mean age 21.44 $\pm$ 2.35 years) with no history of neurological or psychiatric disorders. The total sample comprised 41 males and 59 females. The JME and control groups differed significantly in age (two-sample $t$-test, $p = 0.008$). Age and sex were therefore included as covariates in all group-level analyses.

All participants underwent structural MRI on GE MR750 3T scanners using a 32-channel phased-array head coil. High-resolution T1-weighted images were acquired as part of a standardized epilepsy imaging protocol. Intensity inhomogeneity was corrected using the N4 bias-correction algorithm \cite{tustison.2010}. Cortical reconstruction was performed with the FreeSurfer recon-all pipeline \cite{fischl.2012}, including motion correction, intensity normalization, Talairach alignment, skull stripping, and automated tissue segmentation. Subject-specific cortical meshes were extracted, yielding triangulated surfaces. Surfaces were then mapped to MNI space using FreeSurfer’s surface-based registration \cite{fischl.2012}, establishing vertex-wise correspondence across subjects while preserving intrinsic cortical geometry.

\subsection{Estimating cortical folding pattern}

Let $\mathcal{M}$ denote the closed cortical manifold represented as a triangulated surface mesh. Cortical folding was characterized by the vertexwise mean curvature field $h(x)$, estimated at each mesh vertex $v$. Mean curvature was computed using a local quadratic surface fitting approach based on first-ring neighboring vertices \cite{chung.2003.CVPR,joshi.1995}. 
In local coordinates $(x_1,x_2)$, the surface was approximated by fitting a second--order polynomial using least--squares estimation,
\[
f(x_1,x_2)
=
\beta_0
+
\beta_1 x_1
+
\beta_2 x_2
+
\tfrac{1}{2}\beta_3 x_1^2
+
\beta_4 x_1 x_2
+
\tfrac{1}{2}\beta_5 x_2^2 .
\]
Then the mean curvature is given by \cite{chung.2003.CVPR}
\[
h(v)
=
\frac{1}{2}\,
\frac{(1+\beta_2^2)\beta_3 + (1+\beta_1^2)\beta_5 - 2(\beta_1\beta_2)\beta_4}
{1+\beta_1^2+\beta_2^2}.
\]
By convention, positive curvature corresponds to sulcal fundi and negative curvature to gyral crowns. The resulting scalar field therefore provides a compact and intuitive encoding of local cortical folding polarity.

For subsequent analysis, the mean curvature field $h(v)$ was normalized by subtracting its area--weighted mean using the finite--element mass matrix ${\bf A}$ \cite{chung.2004.ISBI,huang.2020.TMI}, which approximates the surface area element. The total surface area is given by
\[
\mathrm{Area}(\mathcal M)
=
\int_{\mathcal M} d\mu(x)
\;\approx\;
\mathbf{1}^{\top} {\bf A} \mathbf{1}.
\]
Accordingly, $h$ was mean--centered as
\[
h \;\leftarrow\; h \;-\;
\frac{\mathbf{1}^{\top} {\bf A} h}{\mathbf{1}^{\top} {\bf A} \mathbf{1}},
\]
which enforces the zero--mean constraint $\int_{\mathcal M} h\, d\mu = 0$ need for a well--posed Poisson equation on a closed manifold.

\begin{figure}[t]
	\centering
	\includegraphics[width=1\linewidth]{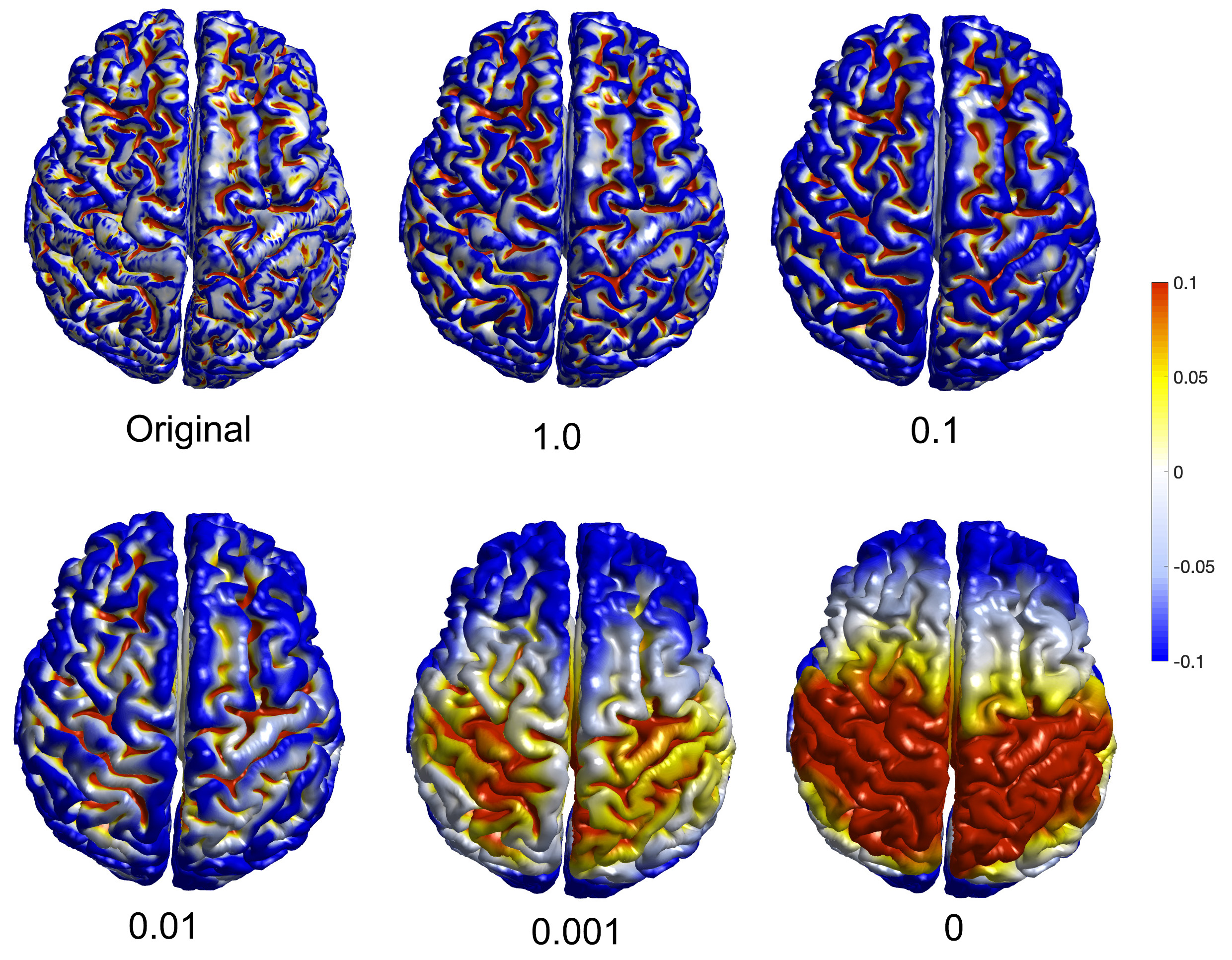}
\caption{Solutions of the regularized Poisson equation for different values of $\lambda$. When $\lambda=0$, the formulation reduces to the standard Poisson equation, which can lead to overly smooth solutions dominated by global modes. In this study, we set $\lambda=0.1$, which balances noise reduction with preservation of local sulcal and gyral pattern.}
\label{fig:poisson}
\end{figure}

Removing the global offset isolates spatially varying folding patterns and removes dependence on an arbitrary baseline curvature. The resulting potential and flow fields therefore capture relative geometric variation between sulcal and gyral regions. By convention, sulcal fundi exhibit positive curvature and gyral crowns negative curvature, so the sign of $h$ encodes local folding polarity (Fig.~\ref{fig:poisson}, top left). Neuronal number per unit cortical area is approximately conserved across cortex \cite{rakic.1988}, thinner sulcal cortex implies higher neuronal packing density per unit volume relative to gyral cortex. The signed curvature field $h$ provides a compact geometric descriptor that is consistent with known microstructural differences between sulcal and gyral regions \cite{vanessen.1997}.

\subsection{Poisson model of cortical folding}

Curvature estimates are sensitive to discretization error, mesh irregularity, and high--frequency geometric noise~\cite{chung.2003.CVPR}. To reduce these effects, we smooth the folding pattern using a Poisson--based model. Let $\mathcal{L}$ denote the Laplace--Beltrami operator on the closed cortical manifold $\mathcal{M}$. Given the area--normalized signed mean curvature field $h$ (+ in sulci and - in gyri), we recover a scalar potential $u$ by solving
\bqn
\mathcal{L}u = h \qquad \text{on } \mathcal{M}.
\label{eq:poisson}
\eqn
The inverse operator $\mathcal{L}^{-1}$ acts as a low--pass filter, suppressing high--frequency noise while emphasizing large--scale source--sink structure in $h$. On a closed manifold, however, global modes may dominate the solution  (Fig. \ref{fig:poisson}, bottom right). To improve spatial locality, we adopt a regularized Poisson model~\cite{kazhdan.2010},
\bqn
\mathcal{L}u + \lambda u = h,
\label{eq:regualized}
\eqn
where $\lambda>0$ controls the balance between smoothness and locality. Small $\lambda$ recovers standard Poisson behavior, while larger values emphasize mesoscale folding structure. We set $\lambda=0.1$ in all experiments.

\begin{figure}[t]
	\centering
	\includegraphics[width=1.0\linewidth]{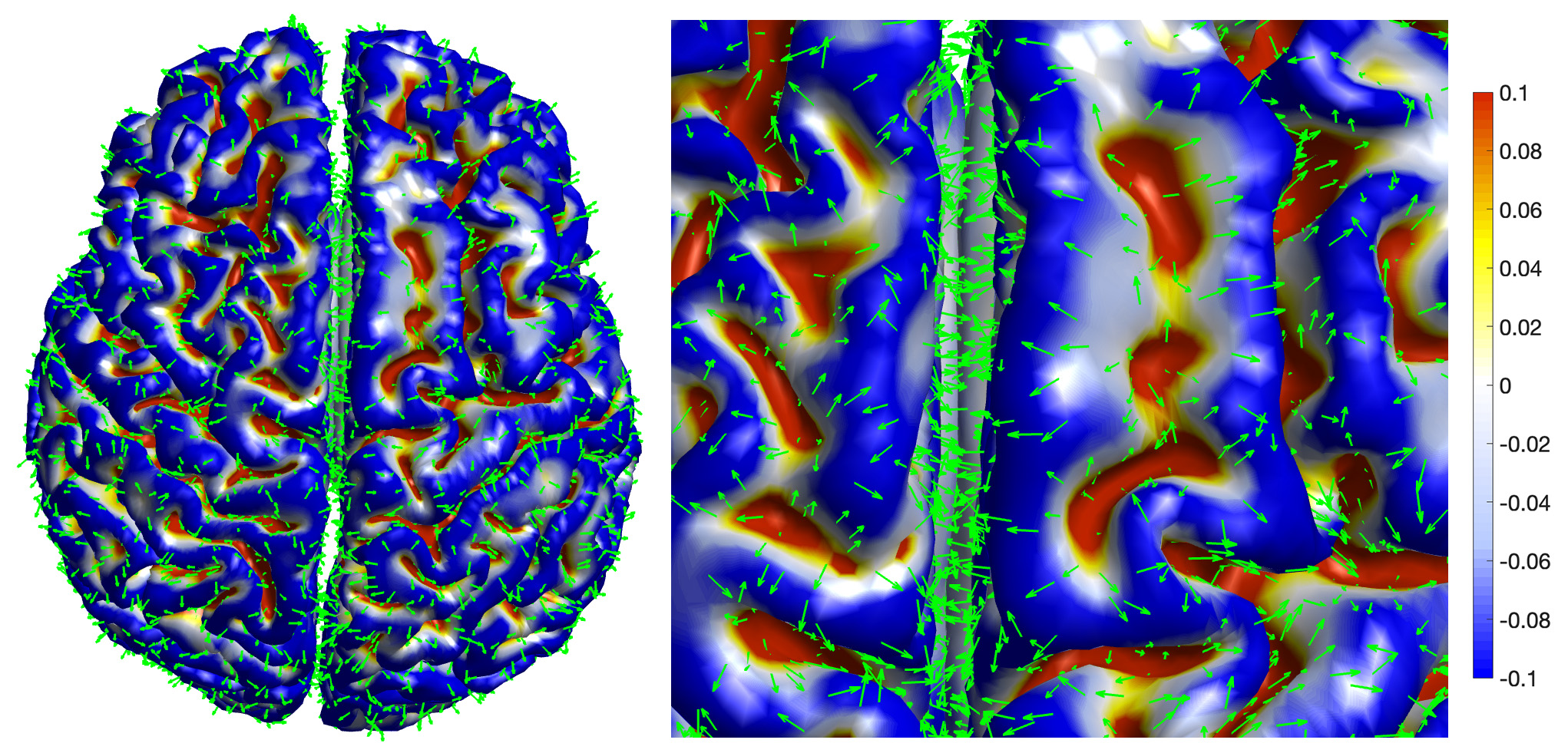}
\caption{\small Flux field defined as the negative surface gradient of the potential, pointing in the direction of steepest decrease of potential energy from sulcal regions (red) toward gyral regions (blue). Compared to the original curvature data, the resulting flux patterns are smoother and  more coherent.}
\label{fig:flux}
\end{figure}

The resulting scalar field $u$ induces a derived vector field $J$ on the cortical surface:
\(
J = -\nabla u,
\)
which lies in the tangent plane and points in the direction of steepest decrease of $u$ (Fig.~\ref{fig:flux}). The gradient field  $J$ encodes spatial variation in the Poisson solution and captures transitions between regions of positive and negative curvature, yielding trajectories that follow local variations from sulcal fundi toward gyral crowns.

\subsection{Numerical implementation}

We solve the regularized Poisson equation using a finite--element discretization \cite{chung.2004.ISBI}. Specifically, we employ the cotangent finite--element formulation of the Laplace--Beltrami operator~\cite{chung.2003.CVPR,huang.2020.TMI}, which yields a sparse, symmetric stiffness matrix ${\bf L}$ together with a finite--element mass matrix ${\bf A}$ that approximates the surface area element. The regularized Poisson equation~(\ref{eq:regualized}) is discretized as~\cite{kazhdan.2010}
\[
({\bf L}+\lambda{\bf A})u = {\bf A}h.
\]
After enforcing the zero--mean constraint on $h$, the resulting linear system is solved using sparse Cholesky factorization~\cite{davis.2006}. For all real data analyses, we fixed $\lambda=0.1$ to balance numerical stability and spatial locality of the recovered folding patterns.  The MATLAB implementation is available at \url{https://github.com/laplcebeltrami/poisson}.

\begin{figure}[t]
	\centering
	\includegraphics[width=1\linewidth]{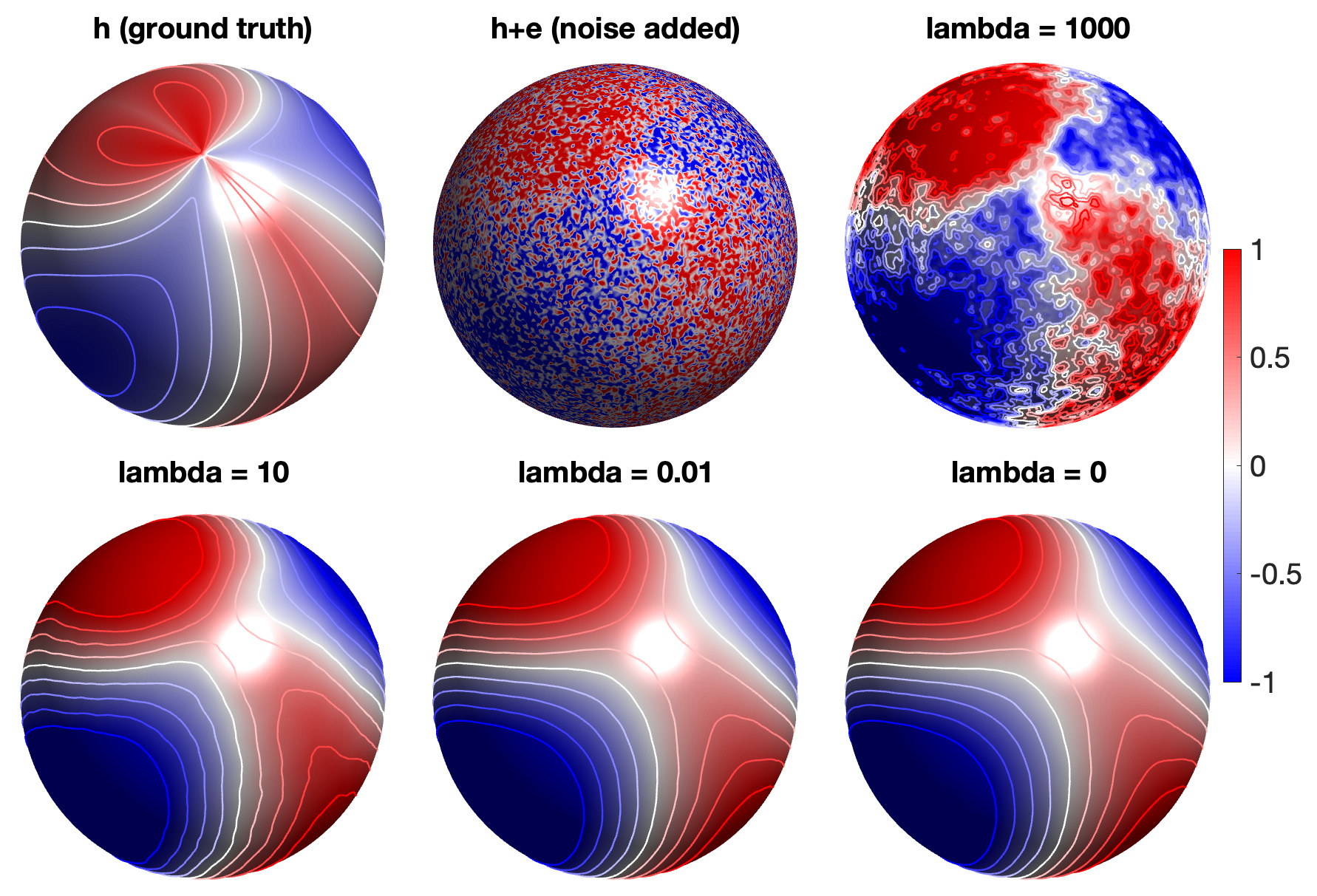}
	\caption{\small
The ground-truth field $h_{\text{true}}$ (top-left) is a degree--2 spherical-harmonic mixture. A noisy observation $h_{\text{obs}}=h_{\text{true}}+\epsilon$ is generated with Gaussian noise $N(0,1)$. The remaining panels show potentials $u$ recovered from the regularized Poisson equation for different $\lambda$. As $\lambda$ decreases, the solution approaches the Poisson limit and becomes more sensitive to high-frequency noise, while larger $\lambda$ suppresses fine-scale fluctuations and emphasizes low-frequency structure.}
\label{fig:validaiton}
\end{figure}

\section{Validation}
\label{sec:validation}

Validation was performed using synthetic data on the unit sphere $\mathbb{S}^2$, where the Laplace--Beltrami operator admits closed--form eigenfunctions given by spherical harmonics $Y_{\ell m}$ \cite{chung.2007.TMI,huang.2020.TMI}. These satisfy
\[
\mathcal{L}Y_{\ell m}=\ell(\ell+1)Y_{\ell m},
\]
providing an analytic setting in which exact solutions of the regularized Poisson equation~(\ref{eq:regualized}) are available. A ground--truth curvature field was constructed as a finite mixture of spherical harmonics,
\[
h_{\mathrm{true}}(\theta,\phi)=\sum_{r=1}^{R}\alpha_r\,Y_{\ell_r m_r}(\theta,\phi),
\]
for which the corresponding regularized Poisson solution is given analytically by
\[
u_{\mathrm{true}}(\theta,\phi)=\sum_{r=1}^{R}\frac{\alpha_r}{\lambda+\ell_r(\ell_r+1)}\,Y_{\ell_r m_r}(\theta,\phi).
\]

To emulate noise arising from discretization and mesh irregularity, Gaussian noise $\epsilon\sim\mathcal{N}(0,\sigma^2 I)$ was added to obtain the observed field $h_{\mathrm{obs}}=h_{\mathrm{true}}+\epsilon$. The recovered potential $u$ was then compared against both $u_{\mathrm{true}}$ and $h_{\mathrm{obs}}$ to quantify numerical accuracy and noise suppression behavior.

In our experiments, we used degree--2 harmonics,
\[
h_{\mathrm{true}}=Y_{20}-Y_{21}+Y_{22},
\qquad
u_{\mathrm{true}}=\frac{1}{6+\lambda}h_{\mathrm{true}},
\]
providing an explicit ground truth for validating our numerical implementation.

Numerical accuracy was quantified using the area--weighted $L^2$ norm $\|f\|^2=f^{\top}\mathbf{A}f$. The error $\|u-u_{\mathrm{true}}\|$ measures solver accuracy, while $\|u-h_{\mathrm{obs}}\|$ reflects regularization behavior relative to noisy input. Results in Table~\ref{table1} show excellent agreement with the analytic solution in the noise-free case and stable performance under noise. Increasing $\lambda$ suppresses noise at the cost of attenuating signal magnitude, demonstrating the expected trade-off between fidelity and smoothing. Overall, the results confirm both the numerical accuracy of the FEM discretization and the noise-filtering behavior of the regularized Poisson model.

\begin{table}[t]
\centering
\caption{\small Performance of the Poisson flow model $u$ on the unit sphere with degree--2 ground truth $u_{\mathrm{true}}$. Errors are reported as area--weighted $L^2$-norms.}
\label{table1}
\setlength{\tabcolsep}{4pt}
\renewcommand{\arraystretch}{1.1}
\small
\begin{tabular}{c c cc}
\hline
Noise $\sigma$ & $\lambda$ & $\|u-u_{\mathrm{true}}\|$ & $\|u-h_{\mathrm{obs}}\|$ \\
\hline
0
 & $10^3$    & $1.66{\times}10^{-5}$ & $1.66{\times}10^{-2}$ \\
 & $10$      & $5.88{\times}10^{-3}$ & $9.41{\times}10^{-2}$ \\
 & $10^{-2}$ & $1.97{\times}10^{-2}$ & $1.19{\times}10^{-1}$ \\
 & $0$       & $1.97{\times}10^{-2}$ & $1.19{\times}10^{-1}$ \\
\hline
0.1
 & $10^3$    & $2.23{\times}10^{-5}$ & $6.57{\times}10^{-2}$ \\
 & $10$      & $5.89{\times}10^{-3}$ & $1.18{\times}10^{-1}$ \\
 & $10^{-2}$ & $1.97{\times}10^{-2}$ & $1.39{\times}10^{-1}$ \\
 & $0$       & $1.97{\times}10^{-2}$ & $1.39{\times}10^{-1}$ \\
\hline
1
 & $10^3$    & $1.46{\times}10^{-4}$ & $6.87{\times}10^{-1}$ \\
 & $10$      & $6.19{\times}10^{-3}$ & $8.04{\times}10^{-1}$ \\
 & $10^{-2}$ & $2.06{\times}10^{-2}$ & $8.13{\times}10^{-1}$ \\
 & $0$       & $2.07{\times}10^{-2}$ & $8.13{\times}10^{-1}$ \\
\hline
\end{tabular}
\end{table}

\section{Results}

Our primary quantity of interest is the folding-induced flow
\(
J = -\nabla u,
\)
which characterizes geometry-driven transport along the cortical surface (Fig.~\ref{fig:flux}). Direct inference on \(J\) is challenging because it is a vector field constrained to the tangent plane. Instead, we exploit the fact that \(J\) is uniquely determined by the scalar potential \(u\), such that testing for group differences in \(J\) is equivalent to testing differences in \(u\). This enables statistically efficient scalar inference.

\begin{figure}[t]
	\centering
	\includegraphics[width=1.0\linewidth]{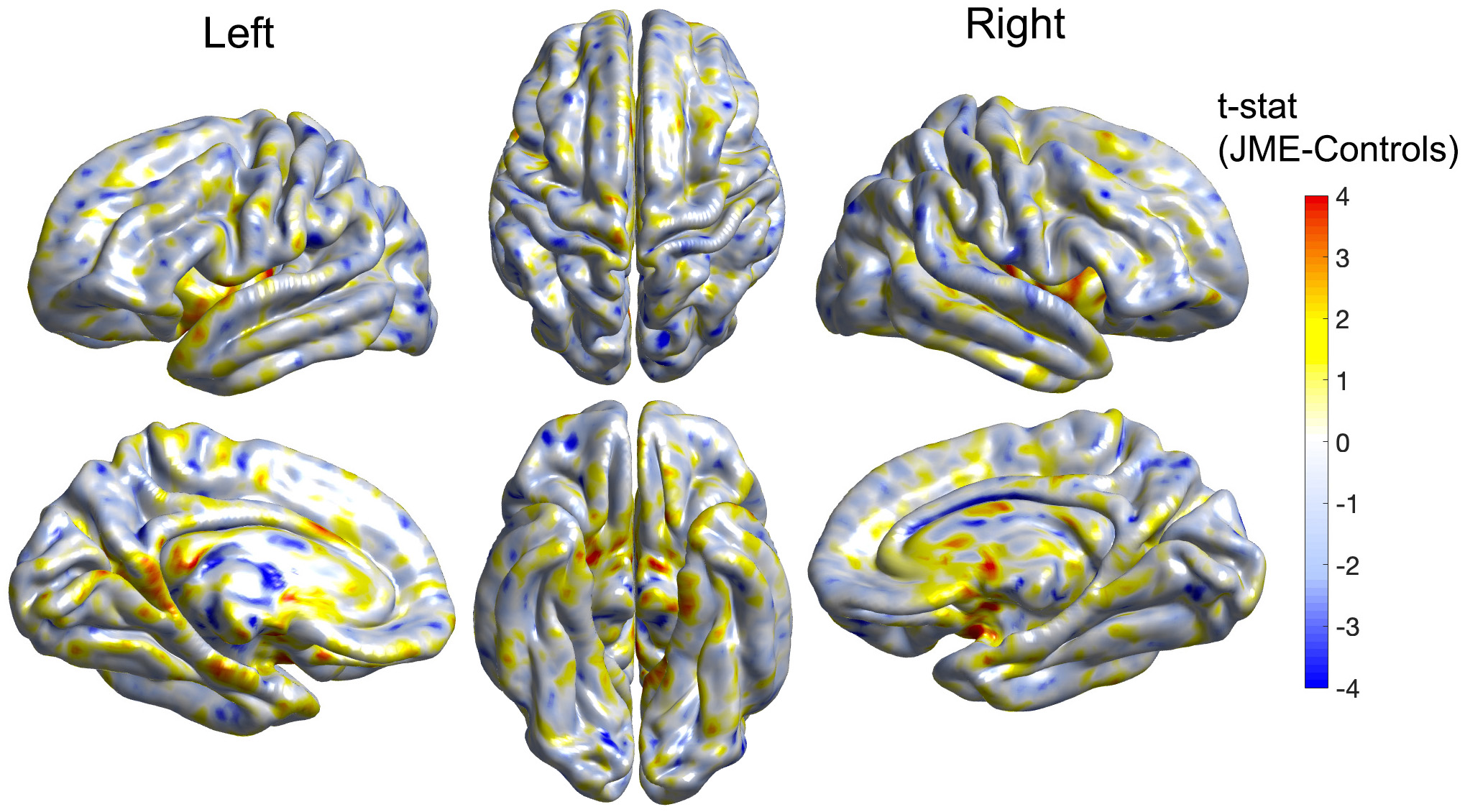}
	\caption{\small $t$-statistic map for the JME--control contrast adjusted for age and sex. Red clusters (JME$>$controls) indicate increased folding potential, with left-dominant effects in fronto-central and medial motor regions and focal right-hemisphere effects along the Sylvian fissure.}
\label{fig:t-stat}
\end{figure}

Group differences in cortical folding were assessed using  linear regression on the folding potential maps \(u_i(v)\). At each vertex \(v\), subject-level potentials were modeled as
\bqn
{\bf u}
=
c_0
+
c_1\,{\tt Group}
+
c_2\,{\tt Age}
+
c_3\,{\tt Sex}
+
\varepsilon,
\label{eq:regression}
\eqn
where ${\tt Group}$ encodes diagnostic status (JME vs.\ control), and age and sex were included as nuisance covariates. Statistical significance of the group effect $c_1$ was evaluated using a  $t$-statistic (Fig.~\ref{fig:t-stat}) with false discovery rate (FDR) correction across the cortical surface \cite{genovese.2002}. Dark red and blue regions ($|t|\geq4.3$) correspond approximately to an FDR-corrected $q$-value of 0.01, while lighter colors ($|t|\geq3.2$) correspond to $q=0.05$.

Regions with significantly elevated folding potential in JME (red clusters) were spatially focal and concentrated along bilateral perisylvian and temporal cortices, with additional involvement of fronto-central and medial motor regions \cite{wandschneider.2019,koepp.2013}. Increased folding potential reflects enhanced curvature contrast and more coherent sulcal–gyral organization, indicating localized alterations in cortical folding geometry. Regions with reduced folding potential in JME (blue clusters) were more widespread and bilaterally distributed, predominantly involving postcentral and visual cortices, with additional focal effects in the right orbital gyri. Reduced folding potential reflects smoother, less differentiated sulcal–gyral geometry, consistent with attenuated large-scale folding organization.

Overall, the observed pattern indicates spatially distributed alterations in cortical folding in JME, characterized by regionally specific accentuation of folding geometry alongside more diffuse reductions in folding differentiation across sensory and visual systems, rather than focal cortical abnormalities~\cite{wandschneider.2019,struck.2025}.

\section{Conclusion}

We introduced a geometry-driven Poisson flow framework for modeling cortical folding organization and applied it to juvenile myoclonic epilepsy (JME). By formulating folding as a curvature-derived source–sink system and solving a regularized Poisson equation on the cortical manifold, the method yields a smooth scalar potential whose gradient defines an intrinsic folding-induced flow. This approach provides a principled link between local sulcal–gyral polarity and large-scale geometric organization, while avoiding the instability and interpretational challenges of direct vector-field inference.

Applied to JME, the results reveal asymmetric and bidirectional alterations in cortical folding organization. Regions of increased and decreased folding potential are spatially distributed across motor, perisylvian, and association cortices rather than confined to a single locus. This pattern indicates spatially heterogeneous structural differences in JME, suggesting that folding alterations are distributed across cortical regions rather than reflecting a focal abnormality~\cite{struck.2025}.

Applied to JME, the results reveal asymmetric and bidirectional alterations in cortical folding organization. Regions of increased and decreased folding potential are spatially distributed across motor, perisylvian, and association cortices rather than confined to a single locus. This pattern indicates spatially heterogeneous structural differences in JME, suggesting that folding alterations are distributed rather than reflecting a focal abnormality~\cite{struck.2025}.

Future work will examine the relationship between folding patterns and clinical variables such as seizure frequency, disease duration, and cognitive performance to assess their clinical relevance. Integrating this framework with functional imaging may further clarify how structural folding organization relates to seizure dynamics. In addition, extensions to multiscale or adaptive regularization and longitudinal analyses may help characterize developmental trajectories underlying the observed group differences.

\bibliographystyle{IEEEbib}

\end{document}